\documentclass[twoside]{article}
\usepackage{fleqn,espcrc2}
\usepackage{graphicx}
\usepackage[figuresright]{rotating}

\newcommand{\AmS}{{\protect\the\textfont2
  A\kern-.1667em\lower.5ex\hbox{M}\kern-.125emS}}
\title{Thermal conductivity in the vortex state of 
$d$-wave superconductors}

\author{I. Vekhter
	\address{Department of Physics, University of Guelph, Guelph,
			Ontario, Canada N1G 2W1}%
        \thanks{Acknowledges the hospitality
	of Centre \'Emile Borel and Aspen Center for Physics}
        and 
        A. Houghton
	\address{Department of Physics, Brown University,
        Providence, RI 02912-1843, USA}}
       
\begin{document}

\begin{abstract}
We present the results of a microscopic calculation of the longitudinal
thermal conductivity of quasiparticles, $\kappa_{xx}$, in a 2D
d-wave superconductor in the vortex state.
Our approach takes into account both impurity scattering
and a contribution to the thermal transport lifetime
due to the scattering of quasiparticles off of vortices. 
We compare the results with the experimental
measurements on high-T$_c$ cuprates and organic superconductors.
\vspace{1pc}
\end{abstract}
\maketitle

In the last few years measurements
of the thermal conductivity have become one of the most
powerful probes of unconventional superconductors. 
Materials with linear nodes in the energy gap have been predicted to
exhibit a residual linear
in the temperature, $T$, term in $\kappa_{xx}$ as $T\rightarrow 0$,
which is only weakly sensitive to
impurity concentration.\cite{lee}
Both properties have been firmly established
in the high-T$_c$ cuprates,\cite{louis}
and a linear term has been resolved in
the organic 
superconductor $\kappa$-(BEDT-TTF)$_2$Cu(NCS)$_2$.\cite{belin}

Information on near-nodal quasiparticles can be obtained
from the analysis of transport properties in the vortex state.
The number of unpaired
quasiparticles increases in an applied magnetic field, $H$,\cite{volovik} 
and their existence leads to
the sublinear, in $H$, increase
of $\kappa_{xx}(H)$ below $T\sim 1$K.\cite{aubin,peter}
In contrast, at higher temperatures the thermal conductivity of the
 cuprates
has been found to decrease
with the applied field, and often becomes nearly field independent above
a few Tesla.\cite{ong} 
Recently we have proposed a unified description
of both features in the cuprates.\cite{vekhter}
In these materials the experiments are restricted to
fields $H\ll H_{c2}$, while
in the organic materials the entire $H-T$
range can be probed, making the analysis of $\kappa_{xx}(H, T)$
important.

We consider a clean 
($l\gg \xi_0$, where $l$ is the mean free path, and
$\xi_0$ is the coherence length) 2-dimensional $d$-wave
superconductor with a cylindrical Fermi surface, model
the vortex lattice by an Abrikosov-like solution,
 and use the quasiclassical
method with the approximation scheme of Ref.\cite{brandt}.
For $l\gg(2eH)^{-1/2}\equiv\Lambda$ 
the thermal conductivity is given by\cite{vekhter}
\begin{eqnarray}
&&{\kappa\over T}={N_0 v^2 \over 4\pi}
\int_0^\infty {d\omega\over T} \biggl({\omega\over T}\biggr)^2
\cosh^{-2}\biggl({\omega\over 2T}\biggr)
\\
\nonumber
&&\qquad\qquad\times
\int_0^{2\pi}d\phi
\cos^2\phi
 \biggl[Re P(\widetilde\omega) \biggr]\tau(\widetilde\omega,\phi),
\\
\label{kernel}
&&{1\over2\tau(\widetilde\omega,\phi)}=
Re\sigma(\widetilde\omega)
\\
\nonumber
&&\qquad\qquad +
2\sqrt\pi{\Lambda\Delta^2\over v} 
{Re [P(\widetilde\omega)W(u)]
\over Re P(\widetilde\omega)}\cos^2 2\phi,
\end{eqnarray}
where $N_0$ is the normal state density of states, $v$ is the Fermi velocity,
  $ P(\widetilde\omega_n)=
[ 1- i \sqrt{\pi} 
                      \bigl({2 \Lambda \Delta /v }\bigr)^2
                           W^{\prime}
                    \bigl (u_n)\cos^2 2\phi]^{-1/2}$,
the function $W(u)=e^{-u^2}{\rm erfc}(-iu)$, 
$u_n= {2 i \widetilde\omega_n \Lambda{\rm sgn}
                                     (\omega_n)
                                     /v}$,
the renormalized frequency
$i\widetilde\omega_n=i\omega_n+i\sigma(i\widetilde\omega_n)$,
depends on the self-energy
due to impurity scattering, $\sigma$,
and $\Delta$ is the spatial average of
the amplitude of the
order parameter.
The second term in the effective scattering rate is due to
scattering off of the vortex lattice. This
term has the same symmetry as the superconducting gap, and therefore
is more important at higher temperatures. At low $T$ and $H$
the main effect of the magnetic field is to
increase the number of the excited quasiparticles near the nodes, where
vortex scattering is weak; consequently the thermal conductivity
increases. At higher fields the number of 
quasiparticles is controlled by the temperature, the main effect
of the vortices is to introduce a new scattering mechanism, and
$\kappa_{xx}$ decreases to a plateau-like feature at fields such that
$v/\Lambda\geq T\gg \sigma(\omega=0)$, in agreement
with the experimental results on high-T$_c$ cuprates.\cite{vekhter}
The parameters of the theory are the dimensionless quantity
$\delta\equiv\Lambda\Delta/v$, and the scattering rate in the normal state, 
$\Gamma$; the scattering phase shift is fixed to either 
the strong (unitarity) or the  weak (Born) limit.
At the mean field level
$\delta=\alpha [(H_{c2}-H)/HH_{c2}]^{1/2}$ where
$\alpha=(\Delta_0/\hbar v)(\hbar c/2e)^{1/2}$. For high-T$_c$ materials 
$\alpha\approx 2$-$7$T$^{1/2}$, \cite{vekhter}
while for $\kappa$-(BEDT-TTF)$_2$Cu(NCS)$_2$
using $\Delta_0=2.4$meV and $v=5\cdot 10^6$cm/s \cite{belin} we find
$\alpha\simeq 1.3$T$^{1/2}$. 

At low $T$ in this material 
$\kappa_{xx}(H)$ decreases sharply below $H_{c2}$.\cite{belin}
In the field range
$H\leq H_{c2}$ at $T=0$ we find from Eqs.(1)-(2)
\begin{equation}
{\kappa(H)\over \kappa_n}\approx {l_{\rm eff}\over l}{1\over \sqrt{1+\mu}}.
\end{equation}
Here $\kappa_n$ is the normal state thermal conductivity, the effective 
mean free path $l_{\rm eff}=v/(2\sigma(\omega=0))$, 
and $\mu=4\sqrt\pi(l_{\rm eff}/\Lambda)\delta^2$.
Near $H_{c2}$ $\mu\propto\Delta^2\propto H_{c2}-H$, so that
$\kappa(H)/\kappa_n\approx 1-\mu/2$ decreases linearly with field.
The range over which
the linear dependence holds, $\mu\ll 1$,
 is smaller the cleaner the sample since for $l\gg\xi_0$
we may have $\mu\geq 1$ even when $\delta\ll 1$. 
Consequently the linear behavior 
down to 0.7$H_{c2}$ \cite{belin} indicates a relatively
dirty sample, in agreement with the  estimate of Ref.\cite{belin}
which yielded
$l\simeq 350$\AA, or $l/\xi_0\simeq 7$.
From this estimate we obtain $\Gamma\simeq 0.4$meV;
$\kappa_{xx}$ determined with these parameters is shown in Fig.1.

\begin{figure}
\includegraphics*[width=7cm]{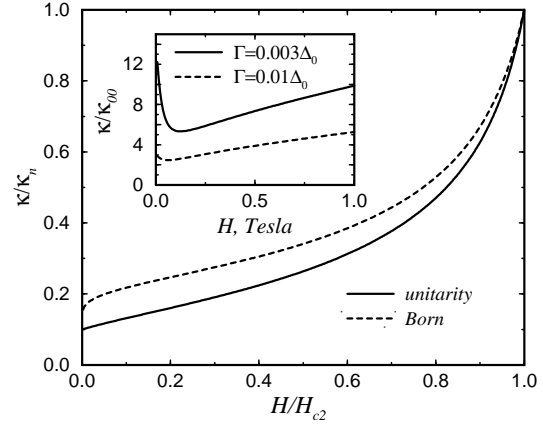}
\caption{Field dependence of $\kappa_{xx}(H,T=0)$ for 
$\Gamma=0.15\Delta_0$, $\alpha=1.3$, and $H_{c2}=5$T.
Inset: $T=0.03\Delta_0$, unitarity limit.
$\kappa_{00}=\pi^2 T N_0 v^2/6\Delta_0$ is the universal 
conductivity.\cite{lee}}
\end{figure}

At low $T$, for fields below $0.5 H_{c2}$,
$\kappa_{xx}(H)$ is observed to decrease with increasing field.\cite{belin}
Note that the residual linear term in 
$\kappa_{xx}(H=0, T)$ is only observable
if $T\leq\sigma(\omega=0)$,
which is exponentially small, in $\Delta_0/\Gamma$, in the Born limit.
The results of Ref.\cite{belin} therefore suggest that the phase shift
is close to $\pi/2$. But, in the unitarity limit 
$\sigma(\omega=0)\approx\sqrt{\Gamma\Delta_0}$, 
and for  $\Gamma\simeq 0.4$meV at low $T$ $\kappa_{xx}(H)$ increases 
rather than decreases over almost the entire field range.
If inelastic scattering is strong,
$\Gamma$ at low $T$ and $H$ may be much smaller than its normal state
value, in which case $\kappa_{xx}$ does 
decrease but 
in  a narrow field range, see Fig. 1,
in disagreement with experiment, suggesting
that the low-field decrease is in the phonon
contribution to $\kappa_{xx}$.

Helpful discussions with K. Behnia and S. Belin are gratefully acknowledged.

\end{document}